\documentclass[12pt,titlepage]{utarticle}
\usepackage{amsmath,amssymb,amsthm,pictexwd,graphicx,color}
\usepackage{setspace}
\usepackage[colorlinks=true, pdfstartview=FitV, linkcolor=blue, citecolor=blue, urlcolor=blue]{hyperref}

\numberwithin{equation}{section}

\def\defeq{\buildrel\rm def\over=}

\def\del{\partial}
\def\wdg{{\wedge}}                              

\def\BZ{\mathbb{Z}}

\newtheorem*{ithm}{Theorem}

\def\Hom{\mathrm{Hom}}
\def\End{\mathrm{End}}

\def\Ext{\mathrm{Ext}}
\def\Coh{\mathrm{Coh}}

\def\mcO{\mathcal{O}}
\def\mcA{\mathcal{A}}

\def\mcC{\mathcal{C}}
\def\mcD{\mathcal{D}}
\def\mcE{\mathcal{E}}
\def\mcF{\mathcal{F}}

\def\Tr{\mathrm{Tr}}

\def\ibar{{\bar{\imath}}}
\def\jbar{{\bar{\jmath}}}

\def\ie{\textit{i.e.,\ }}

\begin{document}
\preprint{MIFP--08--22\\}

\title{Topological D-branes from Descent}

\author{Aaron Bergman\address{George P. \& Cynthia W. Mitchell Institute for Fundamental Physics\\
             Texas A\&M University\\
             College Station, TX 77843-4242\\ {~}\\
             }}

\Abstract{Witten couples the open topological B-model to a holomorphic vector bundle by adding to the boundary of the worldsheet a Wilson loop for an integrable connection on the bundle. Using the descent procedure for boundary vertex operators in this context, I generalize this construction to write a worldsheet coupling for a graded vector bundle with an integrable superconnection. I then compute the open string vertex operators between two such boundaries. A theorem of J. Block gives that this is equivalent to coupling the B-model to an arbitrary object in the derived category.}

\maketitle
\newpage

\section{Introduction}\label{sec:intro}

One of the primary techniques one can use to study nonperturbative aspects of string theory is to place boundary conditions on the string worldsheet that correspond to coupling the string to non-perturbative solitonic objects known as D-branes. While determining all possible boundary conditions in the full string is a daunting proposition, string theory admits certain topological twists that are often simpler to deal with but which still contain significant interesting information. In particular, the topological A- and B-twist respectively capture the symplectic and holomorphic structure of the target Calabi-Yau manifold. Mirror symmetry interchanges the topological A-model on one half of a mirror pair with the B-model on the other. Initially mirror symmetry related certain Hodge theoretic structures related to the A- and B-models. However, Kontsevich later formulated homological mirror symmetry \cite{Kontsevich:1994ho} which postulates a (quasi-)equivalence between certain (A$_\infty$-) categories associated with the A- and B-model. In particular, to the A-model one associates the Fukaya category (or, more properly, some still not known generalization thereof), and to the B-model one associates the bounded derived category of coherent sheaves. It is now understood through work beginning\footnote{Earlier discussions of the derived category in the context of string theory include \cite{Aspinwall:1998he,Sharpe:1999qz}.} with Douglas \cite{Douglas:2000gi} that these categories in fact encode the boundary conditions and open-string states of the respective topological twist. However, explicitly writing down boundary conditions that correspond to a given object in one of the categories has proven elusive.

In this paper, we will focus on the topological B-model and the bounded derived category of coherent sheaves. This latter object has a somewhat formidable reputation in the physics community. We will see that this reputation is hopefully somewhat undeserved. The problem is that in the mathematical literature the construction of the category is done in the context of algebraic geometry, in particular, on schemes. Thus, constructions in sheaf cohomology and the derived category often refer to injective resolutions, and injective sheaves are ungainly things that seem far afield from any physical considerations. However, string theory does not live in the world of algebraic geometry; it sees the analytic structure. Thus, we can use Dolbeault cohomology to do computations in sheaf cohomology and in the derived category. Somewhat more pedantically, we are using the fine resolutions afforded by the Dolbeault complex to replace the injective resolutions often used in algebraic geometry.

In fact, this has already been implemented in string theory in the case of holomorphic vector bundles by Witten \cite{Witten:1992fb}. There, one begins with a connection on a $C^\infty$ vector bundle and places a Wilson line for this connection along the boundary of the worldsheet. BRST invariance of this boundary action implies that the purely anti-holomorphic part of the curvature vanishes, and a standard theorem from differential geometry says that this is equivalent to the existence of a holomorphic structure on the bundle. Witten computes the string states between the given bundles and sees that they are given by the Dolbeault cohomology groups
$$
H^i_{\overline{\del}}(E^\vee \otimes F) \cong \Ext^i(E,F)\ .
$$
Thus we recover sheaf cohomology and Ext groups through integrable connections and the Dolbeault complex.

What will allow us to generalize this is a theorem of Block \cite{Block:2005de} that identifies an object in the derived category with an integrable \textit{superconnection} on a $\BZ$-graded $C^\infty$ vector bundle. In particular, let $E^\bullet$ be such a vector bundle, and let $\nabla$ be a map
$$
\nabla : A^{(0,\bullet)}(E^\bullet) \to A^{(0,\bullet)}(E^\bullet)
$$
of degree one in the combined degree. Here $A^{(p,q)}(E)$ are the differential forms of type $(p,q)$ valued in $E$. We also impose that the map satisfy a Leibniz rule:
$$
\nabla(e\omega) = \nabla(e) \omega + (-1)^e e \bar{\del}\omega\ .
$$
Here $e$ is an arbitrary element in $A^{(0,\bullet)}(E^\bullet)$, and $\omega$ is in $A^{(0,\bullet)}$. This is a $\BZ$-graded anti-holomorphic version of Quillen's superconnection \cite{Quillen:1985sc}. Finally, we impose an integrability condition analogous to the vanishing curvature condition for holomorphic vector bundles:
$$
\nabla \circ \nabla = 0\ .
$$

The theorem of Block tells us that this data is equivalent to that of an object in the derived category. In particular, this means that there is an equivalence of categories between the derived category\footnote{More properly, it is the full subcategory of the bounded derived category of sheaves of $\mcO$-modules consisting of those objects with coherent cohomology sheaves and where $\mcO$ is the sheaf of holomorphic functions. We will ignore this distinction until section \ref{subsec:block}.} and the category whose objects are given by these superconnections and whose morphisms are given by the cohomology defined by a superconnection in the obvious way -- this will be discussed in detail later. Block has a generalization of his construction to generalized complex manifolds that would be interesting to study in the context of string theory, but we will not attempt to pursue that here.

From the point of view of the physics, Block's theorem and the relation to derived categories is not necessary for this paper. What we will do is construct a generalization of Witten's coupling, taking a superconnection and placing a sort of Wilson line for it along the boundary. It is important to note that this in no way represents a proof that the D-brane category for the open B-model is the derived category. Instead, all that is shown is that the derived category is a full subcategory of that D-brane category and that this subcategory is stable under descent, \ie we find no new boundary couplings by deforming by descended boundary vertex operators. The are certainly other things one can place at the boundary. One example is to use Dirichlet boundary conditions, leading to the more traditional notion of a D-brane. It would be interesting to see if a boundary coupling corresponding to a sheaf supported on a submanifold can be seen to correspond to these Dirichlet conditions. 

This coupling and many of the conclusions drawn from it are derived in a different manner in  \cite{Herbst:2008jq} and in \cite{Kapustin:2008sc}.

The plan of this paper is as follows. In section \ref{sec:oldb}, we give a lightning review of the topological B-model and work through Witten's construction of the coupling to holomorphic vector bundles in significant detail. In section \ref{sec:newb}, we compute the boundary descendants of the open string vertex operators in Witten's construction and show how they give rise to a Wilson line for a superconnection. We compute the open string vertex operators in the presence of these new operators and show that they correspond to Block's construction. Finally, in section \ref{sec:derived}, we give a further discussion of Block's theorem, and we relate the results in this paper to prior presentations of the derived category in the physics literature. This section is independent of the rest of the paper, is more mathematical and can be skipped by the uninterested reader.

\section{Vector bundles and the topological B-model}\label{sec:oldb}

\subsection{The closed string}

The canonical reference for the topological twists of the superstring is \cite{Witten:1991zz}. We will be extremely brief here, referring the reader to that reference for all details. Let $\Sigma$ be the string worldsheet and $M$ a Calabi-Yau threefold. After twisting, the field content of the topological B-model can be summarized as follows:
\begin{equation}
\begin{split}
&\phi : \Sigma \to M\ , \\
&\eta \in \Gamma_\Sigma(\phi^*T^{0,1}M)\ , \\
&\theta \in \Gamma_\Sigma(\phi^*T_{1,0}M)\ , \\
&\rho \in A^1_\Sigma(\phi^*T^{1,0}M)\ .
\end{split}
\end{equation}
Here, $\eta$, $\theta$ and $\rho$ are anticommuting fields, and $\rho$ is a worldsheet one-form. When writing in components, we will use middle alphabet letters for spacetime indices and late letters for worldsheet indices.

The topological B-model possesses a nilpotent anticommuting scalar ``BRST" symmetry given as follows:
\begin{equation}
\label{BRST}
\begin{split}
&\delta\phi^i = 0\ , \\
&\delta\phi^\ibar= i \epsilon \eta^\ibar\ , \\
&\delta\eta^\ibar = \delta\theta_i = 0\ , \\
&\delta\rho^i = -\epsilon d\phi^i\ .
\end{split}
\end{equation}

The bulk Lagrangian can be written in the following form:
\begin{equation}
L = it\int_\Sigma \delta V + tW
\end{equation}
where
$$
V = g_{i\jbar}\left(\rho_z^i \del_{\bar{z}}\phi^\jbar + \rho^i_{\bar{z}} \del_z \phi^\jbar \right)\ ,
$$
and
$$
W = \int_\Sigma \left(-\theta_i D\rho^i - \frac{i}{2}R_{i\ibar j\jbar} \rho^i \wdg \rho^j \eta^\ibar \theta_j g^{k\jbar}\right)\ .
$$
It is shown in \cite{Witten:1991zz} that this action only depends on the K\"ahler metric on $M$ up to terms exact in the BRST symmetry and that the $t$-dependence of the theory is essentially trivial. Thus, the theory localizes giving it many of its useful properties.

\subsection{The open string}\label{subsec:openstr}

In this section, we will carefully review Witten's coupling of the boundary to a holomorphic vector bundle. This construction is given in \cite{Witten:1992fb}, and we will fill in many of the details omitted therein as they will be important for the generalization to superconnections. To begin with, we impose the following boundary conditions: the normal derivative to $\phi$ vanishes on the boundary; $\theta$ vanishes on the boundary; and $(\star\rho) |_{\del M} = 0$. These correspond to free boundary conditions. We will return to the question of boundary conditions at the end of this section.

Next, we wish to add Chan-Paton factors, \ie couple the string to a vector bundle on $M$. Let $V$ be a $C^\infty$ vector bundle on $M$. Let $D$ be a map from $A^{(0,p)}(V) \to A^{(0,p+1)}(V)$ that obeys the Leibniz rule. Thus, it can be written locally as
$$
D = \bar{\del} - A_\ibar\ .
$$
We can combine this with a holomorphic exterior derivative to obtain a connection
$$
d - A_\ibar\ .
$$
We add a Wilson line along the boundary for this connection by inserting the following expression into the path integral:
\begin{equation}
\label{wilson}
\Tr\ \mathrm{Pexp}\left(\int_{\del M} A_\ibar \frac{d\phi^\ibar}{dt} dt + i A_{\ibar,j} \eta^\ibar \rho^j\right)\ .
\end{equation}
This should be thought of as the addition of a boundary action. The first term is the usual expression for a Wilson loop in the bundle $V$. The significance of the second term will become clear in the following computation.

We need to verify that this addition respects the BRST symmetry \eqref{BRST}. This is a subtle calculation because the Wilson line is not in spacetime but is in fact a Wilson line for the bundle $\phi^*V$ restricted to the boundary of the worldsheet with connection given by the local one-form $\phi^*(A) + iA_{\ibar,j} \eta^\ibar \rho^j$. However, the BRST variation of $\phi$ is $\phi^\ibar \mapsto \phi^\ibar + i\alpha \eta^\ibar$. Thus, the BRST symmetry \textit{changes the bundle} that our Wilson line lives in. Because the Wilson line is closed with a trace, there is no obstruction to subtracting the two holonomies, but doing a local calculation of the difference is impeded by the fact that we are not allowed to subtract objects in the fibers of two different vector bundles.

To remedy this difficulty, we need a way to identify the fibers of $\phi^*V$ and $(\phi + \delta\phi)^*V$. Thankfully, such an identification is already provided by the covariant derivative on $V$. Put another way, an identification of two adjacent fibers is equivalent to a horizontal vector field on the fiber. Locally, then, it is sufficient to specify a Lie algebra valued function on $\del M$, and the covariant derivative is precisely such a thing. Then, the identification of fibers is given by the exponential map of this Lie algebra element which we can write to first order as
\begin{equation}
\label{buniden}
1 + A_\ibar \delta\phi^\ibar = 1 + i\epsilon A_\ibar \eta^\ibar\ .
\end{equation}
This is precisely an infinitesimal parallel transport in the $\eta^\ibar$ direction.

Now, we wish to use this identification to compute the BRST variation of an infinitesimal part of the Wilson line which we write as
\begin{equation}
\label{infwilson}
P_1 = 1 + A_\ibar \frac{d\phi^\ibar}{dt} dt + i A_{\ibar,j} \eta^\ibar \rho^j\ .
\end{equation}
Since our identification of the fibers is an isomorphism, and the path ordered exponential is a limit of these infinitesimal Wilson lines, if we ensure that their variations vanish, then the variation of the entire Wilson loop also vanishes. It is worth emphasizing that the identification of fibers is a choice, however, and that choice does not change the value of the variation of the Wilson loop. The choice \eqref{buniden} is one that makes the variation of the infinitesimal parallel transport particularly simple. We will see later that for superconnections, a generalization of this choice is needed. The fact that this natural choice exists is not surprising from the mathematics, but there should be a deeper understanding from the point of view of the worldsheet theory. 

The BRST symmetry acts on \eqref{infwilson} as
\begin{eqnarray*}
P_2 &=& 1 + A_\ibar(\phi + \delta\phi) \frac{d(\phi^\ibar + \delta\phi^\ibar)}{dt} dt - i A_{\ibar,j}(\phi + \delta\phi) (\eta^\ibar + \delta\eta^\ibar) (\rho^j + \delta\rho^j)\\
&=&1 + A_\ibar \frac{d\phi^\ibar}{dt} dt + A_{\ibar,j} \eta^\ibar \rho^j + \epsilon \Big(iA_{\ibar,\jbar} \eta^\jbar \frac{d\phi^\ibar}{dt} dt + iA_\ibar \frac{d\eta^\ibar}{dt} dt + A_{\ibar,j\bar{k}} \eta^\ibar \eta^{\bar{k}} \rho^j +iA_{\ibar,j} \eta^\ibar \frac{d\phi^i}{dt}dt \Big)\ .
\end{eqnarray*}
We next conjugate with \eqref{buniden} to find an expression we can subtract from $P_1$:
\begin{equation}
\label{BRSTvar}
\begin{split}
&\hspace{.7in}(1 + i\epsilon A_\ibar \eta^\ibar) P_2 (1 - i\epsilon A_\ibar(\phi(t+dt)) \eta^\ibar(t+dt)) = \\
&(1 + i\epsilon A_\ibar \eta^\ibar) P_2\left(1 - i\epsilon\left(A_\ibar \eta^\ibar + A_{\ibar,j} \frac{d\phi^j}{dt} \eta^\ibar dt + A_{\ibar,\jbar}\frac{d\phi^\jbar}{dt}\eta^\ibar dt + A_\ibar \frac{d\eta^\ibar}{dt} dt \right)\right)\ .\\
\end{split}
\end{equation} 
Expanding to first order in $\epsilon$, we obtain
\begin{eqnarray*}
&1 + A_\ibar \frac{d\phi^\ibar}{dt} dt + A_{\ibar,j} \eta^\ibar \rho^j& \\ &+ \epsilon \Big(
 iA_{\ibar,\jbar} \eta^\jbar \frac{d\phi^\ibar}{dt} dt + iA_\ibar \frac{d\eta^\ibar}{dt} dt + A_{\ibar,j\bar{k}} \eta^\ibar \eta^{\bar{k}} \rho^j + iA_{\ibar,j} \eta^\ibar \frac{d\phi^i}{dt}dt& \\
& + i [A_\ibar,A_\jbar] \eta^\ibar \frac{d\phi^\jbar}{dt}dt - [A_\ibar,A_{\jbar,k}]\eta^\ibar\eta^\jbar \rho^k & \\
& - i A_{\ibar,j} \frac{d\phi^j}{dt} \eta^\ibar dt -i A_{\ibar,\jbar}\frac{d\phi^\jbar}{dt}\eta^\ibar dt -i A_\ibar \frac{d\eta^\ibar}{dt} dt \Big)\ . &
\end{eqnarray*}
Collecting terms, we see that our identification \eqref{buniden} gives us
$$
\delta P = \epsilon\left(i\left(A_{\ibar,\jbar} - A_{\jbar,\ibar} - [A_\ibar,A_\jbar]\right)\eta^\jbar \frac{d\phi^\ibar}{dt}dt +\frac{1}{2}\del_k\left(A_{\ibar,\jbar} - A_{\jbar,\ibar} - [A_\ibar,A_\jbar]\right)\eta^\ibar\eta^\jbar\rho^k\right)\ .
$$
The first term is the type $(0,2)$ part of the curvature, and the second is the derivative of the same. If we had not included the second term in \eqref{wilson}, we would have recovered the usual fact from differential geometry that the parallel transport around an infinitesimal square is given by the curvature of the connection. The addition of the extra term removes the $(1,1)$ part of the curvature from the expression. Thus, to ensure BRST invariance, it suffices to impose that $F^{(0,2)} = 0$. This is equivalent to the statement that $D^2 = 0$ or that the connection is integrable. A theorem in differential geometry (see, for example, \cite{DonKon}) tells us that we can place a holomorphic structure on $V$ such that $D$ is a Dolbeault operator on sections of $V$, and we can compute sheaf cohomology using the associated complex.

Let us now address the question of the boundary conditions. The addition of the boundary action \eqref{wilson} changes the statement of Noether's theorem, and we must modify the boundary conditions to obtain the proper equations of motion.\footnote{I would like to thank Eric Sharpe and Ilarion Melnikov for pointing this out to me.} For a nice discussion of this, see \cite{Melnikov:2003zv}. When \eqref{wilson} is abelian, this is straightforward, but for a non-abelian Wilson loop, it is not clear how to proceed. The problem is that a non-abelian Wilson loop is not a classical object. In fact, it is a partial quantization where the nonabelian degrees of freedom arise from a sigma model into a flag manifold. This is discussed in, for example, section 7.7 of Witten's lectures in volume two of \cite{Deligne:1999qp}. Presumably, one can then use the classical expression for the Wilson loop to determine the correct boundary conditions for the nonabelian Wilson loop. A similar remark applies to the non-abelian boundary coupling derived in the following sections, but I will not attempt to derive the relevant boundary conditions here.

\subsection{Open string vertex operators}\label{subsec:osvo}

As a final exercise, we will determine the open string states between two boundary states by computing the boundary vertex operators. In particular, let $V_1$ and $V_2$ be bundles with integrable connections $D_1$ and $D_2$ which we will write locally as $\bar\del - A^1_\ibar$ and $\bar\del - A^2_\ibar$. We will choose a point $p$ on the boundary and add the following term to the path integral
\begin{equation}
\label{wilinsert}
 \mathrm{Pexp}\left(\int_{\gamma_l} A^1_\ibar \frac{d\phi^\ibar}{dt} dt + i A^1_{\ibar,j} \eta^\ibar \rho^j\right) \mcO_p\,\mathrm{Pexp}\left(\int_{\gamma_r} A^2_\ibar \frac{d\phi^\ibar}{dt} dt + i A^2_{\ibar,j} \eta^\ibar \rho^j\right)
\end{equation}
Here $\gamma_l$ and $\gamma_r$ are boundary components that surround $p$. We assume that something closes the loop, but it is not relevant for our calculation. From this expression, we see that, in addition to being made out of the worldsheet fields, $\mcO_p$ must be valued in $(V_1^\vee \otimes V_2)|_{\phi(p)}$. In particular, this means that we cannot consider $\mcO_p$ independent of the surrounding Wilson lines as the BRST transformation does not act on sections our bundle. 

Before addressing this issue, we will first examine what worldsheet fields we can form vertex operators from. We cannot use $\theta$ because it vanishes and $\rho$ because it is a one-form. Thus, we must restrict to $\phi$ and $\eta$. Since $\eta$ is anti-commuting, we can write such fields as
$$
\alpha(\phi,\eta) = \alpha^0(\phi) + \alpha^1_\ibar(\phi)\eta^\ibar + \alpha^2_{\ibar\jbar}(\phi)\eta^\ibar\eta^\jbar + 
\alpha^3_{\ibar\jbar\bar{k}}(\phi)\eta^\ibar\eta^\jbar\eta^{\bar{k}}\ .
$$
Taking into account that the vertex operators are valued in the fiber of $\phi^*(V_1^\vee \otimes V_2)$, we can identify the space of open string vertex operators with differential forms that have purely anti-holomorphic indices and are valued in said bundle, \ie elements of
$$
A^{(0,\bullet)}(V_1^\vee \otimes V_2)\ .
$$

As with the variation of the Wilson lines, we can use our identification of fibers \eqref{buniden} to compute a local BRST variation. However, each of the Wilson loops has a different identification of fibers. Thus, to subtract $\mcO_p$ and $\mcO_p + \delta\mcO_p$, we must use
\begin{equation}
\label{wilBRST}
(1 + i\epsilon A^1_\ibar \eta^\ibar)(\mcO_p + \delta\mcO_p)(1 - i\epsilon A^2_\ibar \eta^\ibar)
\end{equation}
which, for $\mcO_p$ corresponding to a form $\alpha \in A^{(0,\bullet)}(V_1^\vee \otimes V_2)$, is
\begin{equation*}
\alpha +  i\epsilon\left(\bar\del \alpha +  A^1 \wdg \alpha - (-1)^\alpha \alpha\wdg A^2 \right)
\end{equation*}
where we have identified the expressions $A^{1,2}_\ibar\eta^\ibar$ with the forms $A^{1,2}$, and $(-1)^\alpha$ is the $\BZ_2$-grade of $\alpha$. The latter part of this expression is precisely the antiholomorphic part of the covariant derivative on $V_1^\vee \otimes V_2$ induced by the connections on $V_1$ and $V_2$. Thus, the BRST variation of \eqref{wilinsert} is
$$
 i\epsilon \mathrm{Pexp}\left(\int_{\gamma_l} A^1_\ibar \frac{d\phi^\ibar}{dt} dt + i A^1_{\ibar,j} \eta^\ibar \rho^j\right)\left(D_{V_1^\vee \otimes V_2} \alpha\right)\mathrm{Pexp}\left(\int_{\gamma_r} A^2_\ibar \frac{d\phi^\ibar}{dt} dt + i A^2_{\ibar,j} \eta^\ibar \rho^j\right)\ .
 $$
The covariant derivative $D_{V_1^\vee \otimes V_2}$ is integrable and induces the obvious holomorphic structure and Dolbeault operator on $V_1^\vee \otimes V_2$. As a result, we can identify the BRST cohomology of open string vertex operators with
$$
H^\bullet_{\bar\del}(V_1^\vee \otimes V_2) \cong \Ext^\bullet(V_1,V_2)\ .
$$
This is the same space of states as was computed by a different technique in \cite{Witten:1992fb}.

\section{Superconnections and boundary couplings}\label{sec:newb}

\subsection{Descent with Wilson lines}

In this section, we will derive (or at least motivate) the expression for the coupling to a superconnection by computing the topological descendants of the open string vertex operators explored in the previous section. Before beginning, let us briefly review the usual construction. In a topological quantum field theory, none of the computations should depend on where we have placed our vertex operators. Thus, the insertion of an operator at two different points $p$ and $q$ in the path integral should give the same result. Since we are working in a cohomological field theory, we have that $\mcO_p - \mcO_q$ is BRST exact. Taking the limit as $p$ approaches $q$, we obtain:
$$
d\mcO = \delta\mcO^1
$$
where $\mcO^1$ is a one-form operator. Depending on the dimension of our QFT, we can continue this procedure to obtain a BRST-closed $d$-form operator. The top degree descendent can often be exponentiated and inserted into the Lagrangian to give a deformation of the theory. For the bulk vertex operators in the B-model, this is worked out to some extent in \cite{Witten:1991zz} where one encounters the difficulty that one sometime must add terms proportional to the equations of motion to the descent equations.

We would like to perform this procedure for our boundary operators. Since we are working in one dimension, we only need to descend a single step. However, we are again presented with the issue that one should not consider these vertex operators outside of their surrounding Wilson lines. Since we are interested in exploring operators that we can exponentiate, let us assume that we have a vector bundle $V$ with a connection on it written locally as $\bar{\del}-A_\ibar$ and that the Wilson lines on either side of $\mcO_p$ are for this bundle. We can consider a vertex operator inserted at two locations $\mcO_p$ and $\mcO_q$ as above and subtract them. The only new feature here is the addition of the Wilson line along the boundary from $p$ to $q$. Taking $q = p + dt$ and subtracting, we obtain
$$
\mcO_p \left(1+A_\ibar \frac{d\phi^\ibar}{dt}dt + i A_{\ibar,j} \eta^\ibar \rho^j\right) - 
\left(1 + A_\ibar \frac{d\phi^\ibar}{dt}dt + i A_{\ibar,j} \eta^\ibar \rho^j\right) \left(\mcO_p + d\mcO_p\right)\ .
$$
Thus, the topological descendant is given by
\begin{equation}
\label{topdes}
\delta \mcO^1 = d\mcO_p + [A_\ibar,\mcO_p] \frac{d\phi^\ibar}{dt}dt + i[A_{\ibar,j},\mcO_p]\eta^\ibar \rho^j\ .
\end{equation}
Let us now compute the descendant of the operator $\mcO_p = i \alpha_\ibar \eta^\ibar$. We hope to recover the expression \eqref{wilson}. Substituting into \eqref{topdes}, we obtain
$$
\delta \mcO^1 = i \alpha_{\ibar,\jbar} \frac{d\phi^\jbar}{dt}\eta^\ibar dt + i \alpha_{\ibar,j}\frac{d\phi^j}{dt} \eta^\ibar dt
+ i \alpha_\ibar \frac{d\eta^\ibar}{dt}dt+ i [A_\ibar,\alpha_\jbar] \frac{d\phi^\ibar}{dt}\eta^\jbar dt + [A_{\ibar,j},\alpha_{\bar{k}}] \eta^\ibar\eta^{\bar{k}} \rho^j\ .
$$
We guess that the correct operator is $\mcO^1 = \alpha_\ibar \frac{d\phi^\ibar}{dt} dt + i \alpha_{\ibar,j} \eta^\ibar \rho^j$. To compute the BRST variation, we again use \eqref{wilBRST}, giving:
\begin{equation*}
i \alpha_{\ibar,\jbar} \frac{d\phi^\ibar}{dt}\eta^\jbar dt + i \alpha_\ibar \frac{d\eta^\ibar}{dt}dt + \alpha_{\ibar,j\bar{k}} \eta^\ibar \eta^{\bar{k}} \rho^j + i \alpha_{\ibar,j} \eta^\ibar \frac{d\phi^j}{dt}dt + i[A_\ibar,\alpha_\jbar] \eta^\ibar \frac{d\phi^\jbar}{dt} dt- [A_\ibar,\alpha_{\jbar,k}] \eta^\ibar \eta^\jbar \rho^k\ .
\end{equation*}
The condition that $\mcO_p$ is BRST closed gives
$$
\alpha_{\ibar,\jbar} \eta^\ibar \eta^\jbar - [A_\ibar,\alpha_\jbar]\eta^\ibar\eta^\jbar = 0\ .
$$
Comparing the two expressions, we see that we have guessed correctly, and the relation to \eqref{wilson} is verified.

We will now compute the descendent of something where we do not already know the answer, $\mcO_p = \alpha$ where $\alpha$ is an $\End(V)$ valued function on $M$. Applying \eqref{topdes}, we obtain
$$
\alpha_{,i} \frac{d\phi^i}{dt}dt + \alpha_{,\ibar}\frac{d\phi^\ibar}{dt}dt + [A_\ibar,\alpha] \frac{d\phi^\ibar}{dt}dt + i[A_{\ibar,j},\alpha]\eta^\ibar \rho^j\ .
$$
It is not too hard to see that the BRST variation of $-\alpha_{,i} \rho^i$ is
$$
\delta(-\alpha_{,i} \rho^i) =  -i\alpha_{,i\jbar}\eta^\jbar \rho^i + \alpha_{,i}\frac{d\phi^i}{dt}dt -i [A_\ibar,\alpha_{,j}]\eta^\ibar\rho^j \ .
$$
Since $\delta\alpha = 0$ implies that $\alpha_{,\ibar} + [A_\ibar,\alpha] = 0$, we see that these are equal.

Similar calculations give that the descendant of $-i\alpha_{\ibar\jbar}\eta^\ibar\eta^\jbar$ is
$$
\left(\alpha_{\ibar\jbar} - \alpha_{\jbar\ibar}\right)\eta^\ibar \frac{d\phi^\jbar}{dt} dt + i \alpha_{\ibar\jbar,k} \eta^\ibar \eta^\jbar \rho^k\ ,
$$
and
$i\alpha_{\ibar\jbar\bar{k}}\eta^\ibar\eta^\jbar\eta^{\bar{k}}$ gives
$$
(\alpha_{\ibar\jbar\bar{k}} - \alpha_{\ibar\bar{k}\jbar} + \alpha_{\bar{k}\ibar\jbar}) \eta^\ibar \eta^{\jbar}\frac{d\phi^{\bar{k}}}{dt} dt + i \alpha_{\ibar\jbar\bar{k},l} \eta^\ibar\eta^\jbar\eta^{\bar{k}} \rho^l\ .
$$

\subsection{Coupling to superconnections}

If we wish now to place these descendants into a boundary action, we are presented with a puzzle: some of the descendants are fermionic. We can fix this by declaring some of the $\alpha$s to be fermionic, thus making their descendants bosonic. In particular, this implies that we should generalize from an ordinary vector bundle $V$ to a $\BZ_2$-graded vector bundle. In fact, we should go further. Because the fermion number on the worldsheet is part of the $\BZ$-grading of ghost number, we will work with a $\BZ$-graded vector bundle $V^\bullet$. It is easy to see then that we must assign a grade of 1 to $\alpha$, 0 to $\alpha_\ibar$, $-1$ to $\alpha_{\ibar\jbar}$ and $-2$ to $\alpha_{\ibar\jbar\bar{k}}$. Thus, we have
\begin{eqnarray*}
\alpha& : & V^\bullet \to V^{\bullet+1}\ , \\
\alpha_\ibar &:& V^\bullet \to A^{(0,1)}(V^\bullet)\ ,\\
\alpha_{\ibar\jbar} &:& V^\bullet \to A^{(0,2)}(V^{\bullet-1})\ ,\\
\alpha_{\ibar\jbar\bar{k}} &:& V^\bullet \to A^{(0,3)}(V^{\bullet-2})\ .
\end{eqnarray*}

This looks a lot like the components of a superconnection. Recall from the introduction that, given a graded vector bundle $V^\bullet$, a superconnection is given by a map
$$
\nabla : A^{(0,\bullet)}(V^\bullet) \to A^{(0,\bullet)}(V^\bullet)
$$
which is degree one in the combined degree and which obeys the Leibniz rule
$$
\nabla(e\omega) = \nabla(e) \omega + (-1)^e e \bar{\del}\omega
$$
where $e$ is an arbitrary element in $A^{(0,\bullet)}(V^\bullet)$, and $\omega$ is in $A^{(0,\bullet)}$. The Leibniz rule means that we can locally subtract $\bar{\del}$ to give a differential form. Thus, we can write:
\begin{equation}
\label{superlocal}
\nabla = \bar{\del} - \alpha - \alpha_\ibar - \alpha_{\ibar\jbar} - \alpha_{\ibar\jbar\bar{k}}\ .
\end{equation}
This is analogous to choosing a gauge. The computation of descendents above suggests the following addition to the path integral, analogous to \eqref{wilson}:\footnote{Previous appearences of superconnections in the context of boundary couplings include \cite{Kraus:2000nj,Lazaroiu:2003zi,Takayanagi:2000rz}. I thank K. Hori for pointing out these references.}
\begin{equation}
\label{superwilson}
\Tr\ \mathrm{Pexp}\left(\int_{\del M} \alpha^\eta_\ibar \frac{d\phi^\ibar}{dt}dt + \del_i\alpha^\eta\rho^i\right)
\end{equation}
where
$$
\alpha^\eta_\ibar = \alpha_\ibar + (\alpha_{\jbar\ibar} -\alpha_{\ibar\jbar})\eta^\jbar +(\alpha_{\ibar\jbar\bar{k}} - \alpha_{\jbar\ibar\bar{k}} + \alpha_{\jbar\bar{k}\ibar})\eta^\jbar \eta^{\bar{k}}\ ,
$$
and
$$
\alpha^\eta = \alpha + \alpha_\ibar \eta^\ibar+ \alpha_{\ibar\jbar} \eta^\ibar\eta^\jbar+ \alpha_{\ibar\jbar\bar{k}}\eta^\ibar\eta^\jbar\eta^{\bar{k}}\ .
$$

To compute the BRST-variation of this operator, we again need to choose an identification of the bundles $\phi^*V^\bullet$ and $(\phi+\delta\phi)^*V^\bullet$. As before, this is accomplished by a Lie algebra valued function of $\del M$ (once we have chosen a particular trivialization). However, the choice in the previous section does not lead to any simplification in doing this calculation. Instead, we wish to use an identification related to the superconnection \eqref{superlocal}. Thus, our Lie algebra valued function is $\alpha^\eta$, and our identification is given by:
\begin{equation}
\label{supbuniden}
1 + i\epsilon\alpha^\eta\ .
\end{equation}
Notice that $\epsilon$ is fermionic of ghost number -1, so the expression has total ghost number zero. By restricting to the case where $\alpha^\eta = \alpha_\ibar \eta^\ibar$, we recover the identification \eqref{buniden}.

We can now proceed precisely analogously to section \ref{subsec:openstr}. Let
$$
P_1 = 1 + \alpha^\eta_\ibar \frac{d\phi^\ibar}{dt}dt  + i \alpha^\eta_{,i}\rho^i\ ,
$$
and
\begin{equation*}
P_2 = P_1 + \epsilon\left(i\alpha^\eta_{\ibar,\jbar}\eta^{\jbar}\frac{d\phi^\ibar}{dt}dt +
i\alpha^\eta_\ibar \frac{d\eta^\ibar}{dt}dt  + \alpha^\eta_{,i\jbar}\eta^{\jbar}\rho^i 
+i \alpha^\eta_{,i}  \frac{d\phi^i}{dt}dt\right)\ .
\end{equation*}
We conjugate with \eqref{supbuniden} to obtain:
\begin{eqnarray*}
&\left(1 + i\epsilon\alpha^\eta\right)P_2\left(1 - i\epsilon\alpha^\eta(t + dt)\right) &\\
&=\left(1 + i\epsilon\alpha^\eta\right)P_2 \left(1 - i\epsilon\left(\alpha^\eta  
+ \alpha^\eta_{,i} \frac{d\phi^i}{dt} dt + \alpha^\eta_{,\ibar} \frac{d\phi^\ibar}{dt} dt 
+ \alpha^\eta_\ibar \frac{d\eta^\ibar}{dt}dt\right)\right)&\\
&= P_1 + \epsilon\left(i\left(\alpha^\eta_{\ibar,\jbar}\eta^\jbar - \alpha^\eta_{,\ibar} + [\alpha^\eta,\alpha^\eta_\ibar]\right)\frac{d\phi^\ibar}{dt}dt  + \del_i\left( \alpha^\eta_{,\jbar}\eta^\jbar + (\alpha^\eta)^2\right)\rho^i\right)\ .&
\end{eqnarray*}
The term on the right is the derivative of the curvature of the superconnection, and the term on the left differs from the curvature by a combinatorial factor equal to the `form degree' of each term. The integrability condition $\nabla \circ \nabla = 0$ means that they both vanish. Thus, by Block's theorem \cite{Block:2005de} as explained in the introduction and in section \ref{subsec:block}, this boundary coupling corresponds to an object in the derived category.

\subsection{Open string vertex operators}

Finally, we wish to compute the open string vertex operators as in section \ref{subsec:osvo}. Since this calculation contains no new elements, we will be brief. We now take two graded vector bundles $V_1^\bullet$ and $V_2^\bullet$ with superconnections written locally as $\bar\del - A$ and $\bar\del - B$ where $A$ and $B$ are sums of forms. We define $A^\eta$ and $B^\eta$ as above. The open string vertex operators can be identified with the space:
$$
A^{(0,\bullet)}\left(\left(V_1^\bullet\right)^\vee \otimes V_2^\bullet\right)\ .
$$
We consider this as a singly graded complex by taking the form grading plus the $V_2$ grading minus the $V_1$ grading.

To compute the BRST variation, we continue to use the identification of the previous section \eqref{supbuniden}. Thus, \eqref{wilBRST} becomes
$$
(1 + i\epsilon A^\eta)(\mcO_p + \delta\mcO_p)(1 - i\epsilon B^\eta)\ .
$$
For $\mcO_p = -i\nu^\eta$ for some form $\nu$, the BRST variation is given by
$$
\bar\del \nu^\eta + A^\eta \nu^\eta - (-1)^\nu\nu^\eta B^\eta\ .
$$
This defines an integrable superconnection on $(V_1^\bullet)^\vee \otimes V_2^\bullet$, and we see that the BRST cohomology of open string vertex operators is given by the cohomology of this superconnection on $A^{(0,\bullet)}\left(\left(V_1^\bullet\right)^\vee \otimes V_2^\bullet\right)$. This is precisely as one wants from Block's theorem. Thus, we have defined boundary states that correspond to objects in the derived category and open string vertex operators that correspond to morphisms in that category. However, everything we have done so far is completely independent from Block's theorem; the physics has no need to know that these superconnections are related to derived categories. Nonetheless, the derived category already has a history in the topological string, and we will devote the remainder of this paper to a discussion of the mathematical context of these results and their relation to some previous constructions in the physics literature.

\section{Background and context}\label{sec:derived}

\subsection{Block's theorem}\label{subsec:block}

Having gone this far without an explicit statement of Block's theorem, we now remedy that oversight. Let $M$ be a complex manifold. We define the following category, $\mcC$. The objects are given by graded vector bundles with integrable superconnections. There is a shift operator given by shifting the grading of the vector bundle and superconnection. Given two objects $\mathbb{V}_1 = (V_1^\bullet,\nabla_1)$ and $\mathbb{V}_2 = (V_2^\bullet,\nabla_2)$, we define the complex
$$
\Hom^\bullet_\mcC(\mathbb{V}_1,\mathbb{V}_2) = A^{(0,\bullet)}\left(\left(V_1^\bullet\right)^\vee \otimes V_2^\bullet\right)
$$
with a differential given as follows. For $\phi \in A^{(0,\bullet)}\left(\left(V_1^\bullet\right)^\vee \otimes V_2^\bullet\right)$,
$$
(d\phi)(v) =  \nabla_2\phi(v) - (-1)^\phi \phi(\nabla_1v)\ .
$$
That each of the morphism spaces is a complex makes this a differential graded (dg) category. One can associate a category whose morphisms are given by the cohomology of these complexes. This is called the homotopy category of the dg-category.

Block shows that, to any complex of sheaves $\mcF^\bullet$ with coherent cohomology, there corresponds a graded vector bundle and integrable superconnection $\mathbb{F} = (F^\bullet,\nabla)$ such that for any two such complexes, $\mcE_1^\bullet$ and $\mcE_2^\bullet$, we have
$$
\Ext^k(\mcE_1^\bullet,\mcE_2^\bullet) \cong H^k\Hom^\bullet(\mathbb{E}_1,\mathbb{E}_2)\ .
$$

I will not give the details of the construction of a graded vector bundle and superconnection out of a complex of sheaves except to note that the graded vector bundle arises as a resolution of the complex of  sheaves in terms of $C^\infty$ vector bundles, and the degree zero part of the superconnection are the maps in the resulting complex. One then makes use of the fact that vector bundles correspond to projective modules to construct the higher terms in the superconnection and eventually obtain something that squares to zero.

To go the other direction, given a graded vector bundle and superconnection $(E^\bullet,\nabla)$, examine the sheaf of forms of purely antiholomorphic type valued in $E^\bullet$, $\mcA^{(0,\bullet)}(E^\bullet)$.
This can be made into a complex by acting with the superconnection as it is of degree one in the total degree, and it squares to zero. Block shows that the cohomology sheaves of this complex are coherent, and hence \cite{Block:2005de}
\begin{ithm} 
The homotopy category of the category $\mcC$ defined above is equivalent to the full subcategory of the bounded derived category of sheaves of $\mcO$-modules with coherent cohomology sheaves.
\end{ithm}
There is an important subtlety here. There are two possibly distinct categories that we have so far conflated in this paper. One is the bounded derived category of coherent sheaves. This is the category formed by taking bounded complexes of coherent sheaves and inverting quasi-isomorphisms. The second category is the one mentioned in the theorem. To obtain this category, one starts with the bounded derived category of sheaves of $\mcO$-modules and restricts to the objects whose cohomology sheaves are coherent. In algebraic geometry, if our scheme is noetherian, quasicompact and separated, then the categories are equivalent. However, as emphasized in the introduction, physics is not about schemes. Still, if our complex manifold is projective, then by the GAGA theorems we can make use of algebraic techniques and the categories are equivalent. In fact, it is not too hard to see that all Calabi-Yau 3-folds with finite fundamental group are projective\footnote{I thank Dave Morrison for pointing this out to me.}. However, this discussion suggests that for non-projective targets such as certain K3s, the category of the theorem is the appropriate one for physics. As far as I know, the relationship between these two categories is not understood outside of the algebraic situation.

\subsection{Quasi-isomorphisms}

Having made it this far, the reader familiar with common presentations of the derived category may be asking themselves, ``Where have you inverted the quasi-isomorphisms?". As background, let us review the classical construction of the derived category. One begins with a category whose objects are bounded complexes of coherent sheaves and whose morphisms are given by chain maps between those complexes. A chain map is called a \textit{quasi-isomorphism} if it induces an isomorphism on the cohomology sheaves of the complex. The process of inverting quasi-isomorphisms is to add sufficient morphisms to the category such that every quasi-isomorphism has an inverse\footnote{One does not have to quotient by homotopic maps when stated in this generality, but certain specific constructions require it.}. There is a useful alternate way of viewing this. An object is called acyclic if all its cohomology sheaves are zero. One can look at the full subcategory of acyclic objects\footnote{This is an example of a ``thick" subcategory. See, for example, \cite{GelMan}.} and form the quotient of our original category by this subcategory. The resulting quotient makes all acyclic objects isomorphic to the zero object. It is an easy exercise in the axioms of a triangulated category to see that this is equivalent to inverting the quasi-isomorphisms. The resulting category is called the bounded derived category of coherent sheaves, and is denoted $\mcD^b(\Coh(M))$.

As an interesting aside, there is a dg-category whose homotopy category is $\mcD^b(\Coh(M))$. To construct it, one again begins with bounded complexes of coherent sheaves as objects. Given two such complexes $\mcE^\bullet$ and $\mcF^\bullet$, the morphism complex is
$$
\Hom^n(\mcE^\bullet,\mcF^\bullet) \defeq \bigoplus_{i-j=n} \Hom_M(\mcE^i,\mcF^j)
$$
with differential given by $d\phi = \delta_\mcF\phi - (-1)^\phi \phi\delta_\mcE$ where the $\delta$ are the maps in the complexes $\mcE^\bullet$ and $\mcF^\bullet$. The cohomology of this complex is the chain maps modulo homotopy. To invert quasi-isomorphisms, we quotient by the full subcategory of acyclic objects. One way to do so is in Drinfeld \cite{Drinfeld:2004dq} where loosely speaking one adds new morphisms to the category such that the identity morphism of an acyclic object is exact.  This dg-category should be the relevant one for string theory, and the Massey products (or, equivalently, the A$_\infty$ enhancement of the homotopy category) should encode the open string disc amplitudes \cite{Aspinwall:2004bs}. It would be nice to understand the relation between this dg-category and the dg-category arising from superconnections presented above which seems more natural from the worldsheet point of view.

This procedure of inverting quasi-isomorphisms has long seemed mysterious from the point of view of physical constructions with it often being ascribed to such things as RG flow. Thus, one should ask what is the relevance of this procedure to the constructions in this paper. As alluded to in the introduction, what we have done is to avoid the question by working with fine sheaves. To be more specific, it is a standard theorem in homological algebra (\cite[Theorem 10.4.8]{Weibel} for example) that, provided our original abelian category has enough injectives (every object has an injective map into an injective object), the derived category of bounded below complexes is equivalent to the homotopy category of bounded below complexes of injectives. A similar statement holds for bounded above categories when we have enough projectives (every object has a surjective map from a projective). This is how one often does computations in the derived category: one replaces the object at hand with a complex of projectives or injectives (or perhaps some other adapted object) and applies the functor to that complex. What the construction in this paper does is replace our complex of sheaves with a `twisted complex' of fine sheaves. Fine sheaves have no sheaf cohomology (see, for example, \cite{GriHar}) and thus form suitable resolutions for our purposes. If this discussion is too abstract, Block works through an example of how a quasi-isomorphism is inverted following Proposition 2.22 in \cite{Block:2005de}.

Still, one is left with the question about the physical import of having multiple equivalent boundary couplings. Translated into physical terms, we have two seemingly distinct Wilson lines such that there exists open string vertex operators $\mcO_{AB}$ and $\mcO_{BA}$ from $A$ to $B$ and $B$ to $A$ respectively which annihilate when brought close to each other. In particular, a Wilson line of $B$ inside a Wilson line of $A$ can be shrunk to zero without affecting any physical computations, and vice versa. If we take a functionalist point of view, everything one can compute using boundary coupling $A$ is the same as what one computes from boundary coupling $B$, so we may as well consider the boundary couplings as being the same. However, it would be nice to see if the boundary couplings can be thought of as differing by a BRST-exact expression. In particular, this should relate to the question of gauge invariance mentioned earlier.

\subsection{Previous constructions}

To conclude, we will briefly discuss the relation to two previous constructions, those of Aspinwall and Lawrence \cite{Aspinwall:2001pu} and of Diaconescu \cite{Diaconescu:2001ze}. Both constructions rely the existence of resolutions of coherent sheaves by bounded complexes of \textit{holomorphic} vector bundles. This is a statement that holds in algebraic geometry on smooth varieties, but it does \textit{not} hold on general complex manifolds. However, as mentioned above, by GAGA (see also \cite{GriHar}) it also holds on smooth \textit{projective} complex manifolds. While neither paper constructs explicit boundary actions, we can see how such resolutions arise from a superconnection. In particular, let $V^\bullet$ be a bounded complex of holomorphic vector bundles with maps $\beta: V^i \to V^{i + 1}$ such that $\beta_{i+1}\beta_i = 0$. Furthermore, we can consider each $V^i$ as a $C^\infty$ vector bundle with an integrable anti-holomorphic connection $D_i : \Gamma(V^i) \to A^{(0,1)}(V^i)$. As the $\alpha_i$ are holomorphic maps, it is easy to see that the sum $\nabla = \sum_i \beta_i + D_i$ is an integrable superconnection and that the complex of morphisms is precisely the Dolbeault complex for the complexes of holomorphic vector bundles. Thus the construction here both encompasses and extends the constructions in those papers.

\section*{Acknowledgments}

I would like to thank Jonathan Block for sharing a draft of \cite{Block:2005de} and for discussions in understanding his theorem. I would also like to thank Jacques Distler and Ronen Plesser for helpful discussions. While this work was being completed, I enjoyed the hospitality of the Simons Workshop on Mathematics and Physics and the KITP Miniprogram on Gauge Theory and Langlands Duality. I thank them for their support. This research was supported in part by DARPA and AFOSR through the grant FA9550-07-1-0543, by the National Science Foundation under Grants No. PHY05-51164 and PHY-0505757, and by Texas A\&M University.

\bibliographystyle{utphys}
\bibliography{thebib}

\end{document}